\documentclass[pre,twocolumn,aps,superscriptaddress]{revtex4-1}
\usepackage[pdftex, colorlinks=true,
pdfstartview=FitV,
linkcolor= linkcolor,
citecolor= linkcolor,
urlcolor= linkcolor,
hyperindex=true,
hyperfigures=false]
{hyperref}

\usepackage[dvips]{graphicx}
\usepackage{amsmath, mathrsfs, stmaryrd}
\usepackage{color}

\definecolor{linkcolor}{rgb}{0,0,0.6}

\newcommand{\by}{\text{\bf y}}

\newcommand{\cA}{S}

\newcommand{\cO}{\mathcal{O}}
\newcommand{\cW}{p}
\newcommand{\R}{\text{r}}

\newcommand{\beeta}{\boldsymbol{\eta}}

\newcommand{\dd}{\text{d}}

\newcommand{\ee}{\text{e}}
\newcommand{\ii}{\text{i}}

\newcommand{\p}{\partial}
\newcommand{\bx}{\text{\bf x}}

\newcommand{\br}{\text{\bf r}}

\newcommand{\eps}{\varepsilon}
\newcommand{\bk}{\text{\bf k}}

\newcommand{\bq}{\text{\bf q}}

\newcommand{\bv}{\text{\bf v}}

\newcommand{\bll}{{\boldsymbol{\ell}}}

\providecommand{\avg}[1]{\left \langle #1 \right \rangle}
\providecommand{\pnt}[1]{\left ( #1 \right)}
\providecommand{\brt}[1]{\left [ #1 \right]}
\providecommand{\cur}[1]{\left\{ #1 \right\}}

\providecommand{\f}[2]{\frac{#1}{#2}}
\providecommand{\df}[2]{\dfrac{#1}{#2}}

\begin{document}

\title{Non-Gaussian noise without memory in active matter}

\author{\'Etienne Fodor}
\affiliation{DAMTP, Centre for Mathematical Sciences, University of Cambridge, Wilberforce Road, Cambridge CB3 0WA, UK}
\affiliation{Laboratoire Mati\`ere et Syst\`emes Complexes, UMR 7057 CNRS/P7,
Universit\'e Paris Diderot, 10 rue Alice Domon et L\'eonie
Duquet, 75205 Paris cedex 13, France}
\affiliation{Yukawa Institute for Theoretical Physics, Kyoto University,
Kitashirakawa-oiwake cho, Sakyo-ku, Kyoto 606-8502, Japan}

\author{Hisao Hayakawa}
\affiliation{Yukawa Institute for Theoretical Physics, Kyoto University,
Kitashirakawa-oiwake cho, Sakyo-ku, Kyoto 606-8502, Japan}

\author{Julien Tailleur}
\affiliation{Laboratoire Mati\`ere et Syst\`emes Complexes, UMR 7057 CNRS/P7,
Universit\'e Paris Diderot, 10 rue Alice Domon et L\'eonie
Duquet, 75205 Paris cedex 13, France}

\author{Fr\'ed\'eric van Wijland}
\affiliation{Laboratoire Mati\`ere et Syst\`emes Complexes, UMR 7057 CNRS/P7,
Universit\'e Paris Diderot, 10 rue Alice Domon et L\'eonie
Duquet, 75205 Paris cedex 13, France}
\affiliation{Yukawa Institute for Theoretical Physics, Kyoto University,
Kitashirakawa-oiwake cho, Sakyo-ku, Kyoto 606-8502, Japan}

\begin{abstract}
  Modeling the dynamics of an individual active particle invariably
  involves an isotropic noisy self-propulsion component, in the form
  of run-and-tumble motion or variations around it. This
  nonequilibrium source of noise is neither white---there is
  persistence---nor Gaussian. While emerging collective behavior in
  active matter has hitherto been attributed to the persistent
  ingredient, we focus on the non-Gaussian ingredient of
  self-propulsion. We show that by itself, that is without invoking
  any memory effect, it is able to generate particle accumulation
  close to boundaries and effective attraction between otherwise
  repulsive particles, a mechanism which generically leads to
  motility-induced phase separation in active matter.
\end{abstract}

\maketitle

% ===============================================================================

\section{Introduction}

The Brownian dynamics of particles interacting via conservative forces
inevitably leads, whatever the level of friction, to a steady-state
distribution given by the celebrated Gibbs-Boltzmann distribution. The
key property allowing this statement to be made without having to
solve for the dynamics is detailed balance, a signature of time
reversibility. Granted, whatever the specifics of time-reversal, it is
synonymous for equilibrium behavior, and it comes hand-in-hand with a
number of well-known consequences, such as the fluctuation-dissipation
theorem (FDT)~\cite{Kubo1966} or a vanishing entropy production. In active matter, by
contrast, the interest goes to particles whose individual motion,
while isotropic, relies on a net dissipation of energy. Such systems
leave the realm of equilibrium physics, and the door is open to a
wealth of phenomena that our equilibrium intuition often fails to
grasp~\cite{Vicsek:1995:PRL,PhysRevLett.100.218103, Schaller:2010:Nature, Fily:12, Marchetti:2013:RMP,
  Cates:15,Bricard:2013:Nature,Wensink:2012:PNAS,sumino2012large,Palacci:2013:Science}.

Focusing on the subclass of active systems made of isotropic
particles, a variety of microscopic dynamics have been proposed to
model individual motion. Run-and-Tumble Particles (RTPs), for which
directed motion is interspersed by random directional changes, are
amongst the most studied of models and have been used to model
swimming bacteria~\cite{berg2008coli,Schnitzer1993}. Active Brownian Particles
(ABPs), in which a Gaussian white noise drives directional diffusion
at an otherwise constant tangential velocity, provide a simple model
for self-propelled colloids~\cite{Palacci:2010:PRL,Bocquet:12, Fily:12,
  Buttinoni:2013:PRL,Palacci:2013:Science,Ginot:2015:PRX, Suma2018}. In such models,
instead of a standard equilibrium Gaussian white noise mimicking the
action of the solvent on the particles, one has to deal with a random
force that is neither Gaussian nor white.

In an effort to further simplify such active particle models, without
however giving up the gist of nonequilibrium activity, it has been
argued that the main nonequilibrium ingredient was the existence of
some memory, also termed \textit{persistence}, in the random self-propulsion force. This
has led, for instance, to a series of works on Active
Ornstein-Uhlenbeck Particles (AOUPs)~\cite{PhysRevE.91.042310, PhysRevE.91.062304, Maggi:15, PhysRevLett.117.038103, Wittmann2017a, Wittmann2017b,Mandal2017},
in which a Gaussian noise characterized by an exponentially decaying
memory kernel is used. See also~\cite{Levis:2014} for a kinetic Monte Carlo version of AOUPs. In the latter case, of course, no matching
memory kernel in the viscous damping is introduced, else one would
fall back onto equilibrium physics as described by
Kubo~\cite{0034-4885-29-1-306} in his works on generalized Langevin
equations
\footnote{Unless, of course, there is some imbalance between force
  correlations and viscous damping characterized by different memory
  kernels~\cite{berthier:2013:NP}}. AOUPs have been used,
for instance, to model the dynamics of tracers in living
systems~\cite{Gov:11, Visco:15, Ahmed:15c}.

Our purpose in this work is to investigate what physical
characteristics the non-Gaussian nature of the active fluctuations
brings in. We will thus take the opposite stance and forget about any
type of memory, thereby working with a non-Gaussian but white noise in
our equations of motion for the individual particles. In practice we
consider particles experiencing a viscous drag, a random force, and
either external or interparticle forces:
\begin{equation}\label{LangevinNGintro}
  m\frac{\dd \bv_i}{\dd t}=-\gamma \bv_i+ {\bf F}_{i}+\gamma\beeta_i.
\end{equation}
Here, $m$ is the mass of the particles, which we send to zero to
describe an overdamped limit, but retain finite in our simulations for
practical purposes explained below.  We consider a non-Gaussian white
noise $\beeta$, known as a filtered Poisson process with Dirac kernel:
Over a given time interval of duration $t_\text{obs}$, a number $n$ of
times $\{t_1,\ldots,t_n\}$ is drawn at random from a Poisson distribution
with average $\nu t_\text{obs}$. These times are
themselves random variables drawn from a uniform distribution over
$[0,t_\text{obs}]$. At each time $t_i$ a random vector ${\bll}_i$ is
independently drawn from some specified jump distribution $p({\bll})$
so that:
\begin{equation}\label{eq:defnoise}
	\beeta(t)=\sum_{i=1}^n {\bll}_i\delta(t-t_i).
\end{equation}
Interestingly, our main message is that such non-Gaussian dynamics
exhibit much of the standard active matter behaviors frequently
associated with persistent noises, such as accumulation close to
boundaries.

Langevin equations driven by a non-Gaussian white noise have 
been considered before. An early instance can be found in signal
processing~\cite{ROBERTS197223}, but other developments have been
witnessed in mechanical and structural engineering~\cite{grigoriu1987,
  DiPaola1992, dipaolastochastic1993, DIPAOLA1993197, GRIGORIU1995171,
  Pirrotta2007127}, including processes involving mutiplicative
noise, or even more recently in
finance~\cite{popescu2015kramers, Kanazawa2018}. From a physics
perspective, some properties of the harmonic oscillator evolving under
a non-Gaussian white noise have been solved
exactly~\cite{0305-4470-30-24-009, dubkov2005generalized}. More
complex force fields or noises have been investigated in earlier
years~\cite{GRIGORIU1995171,Duarte2011,Duarte2016}. More recently, it
was shown how to properly formulate the ideas of stochastic
thermodynamics in the presence of non-Gaussian white
noise~\cite{PhysRevLett.108.210601} and actual physical realizations
have been brought forth~\cite{PhysRevLett.114.090601, Kanazawa2015,
  Sano2016, kanazawa2017statistical}. These few references do not by any means make up an
exhaustive review. An important feature that is absent from all these
works, however, is that no collective effects (between various
particles evolving with such modified Langevin dynamics) are
considered.

Before we consider interacting particle systems, we first discuss in
Section~\ref{sec:individual} the properties of the microscopic
dynamics that we endow our particles with. The specifics of a
non-Gaussian but white noise are described there. We focus,
analytically and numerically, on a single particle evolving in an
external potential, considering in particular a particle confined in a
harmonic trap and a particle in the vicinity of a wall. This study of
one-body problems is designed to lay the ground for the many-particle
case that we consider in Section~\ref{sec:collective}. We first
establish that, as for persistent active particles, quorum-sensing
interactions that make motility decrease at high density leads to a
motility-induced phase separation
(MIPS)~\cite{PhysRevLett.100.218103,Tailleur:13,Cates:15}. Simulations
of large bidimensional systems of non-Gaussian particles interacting
via pairwise forces are numerically beyond what we can achieve and we
thus could not establish MIPS in this
case~\cite{Fily:12,Hagan:13,Buttinoni:2013:PRL}. We nevertheless
show that, as for ABPs, RTPs or AOUPs, purely repulsive forces
induce an effective attraction between the particles. To interpret
what we observe, we describe the dynamics of collective modes and we
build up an evolution equation for the local particle density {\it \`a
  la} Dean-Kawasaki~\cite{dean_langevin_1996, Kawasaki}.
We use that equation to construct the BBGKY
hierarchy of correlations~\cite{Hansen2013}, and we derive a low-density expansion of
correlation functions.

% ===============================================================================

\section{Single particle dynamics}\label{sec:individual}

\subsection{A modified Langevin dynamics}\label{sec:lang}

In order to pose our problem with care, we begin with the
single-particle version of dynamics~\eqref{LangevinNGintro}:
\begin{equation}\label{LangevinNGunder}
	m\frac{\dd \bv}{\dd t}=-\gamma \bv+{\bf F}+\gamma\beeta ,
\end{equation}
where ${\bf F}=-\p_\br U$ is taken to be a conservative force deriving
from the potential $U$. The factor $\gamma$ in front of $\beeta$ is
here for practical reasons, as we prefer thinking of the noise
$\beeta$ as a fluctuating velocity imparted to the particle.

Exploiting the definition of the non-Gaussian noise $\beeta$ in
Eq.~\eqref{eq:defnoise}, the generic average brackets
$\langle\cdot\rangle$, referring to an average over the noise
realizations, thus denote an average with respect to $n$, to the
$t_i$'s and to the ${\bll}_i$'s. The generating functional of $\beeta$
is
\begin{equation}
	Z[{\bf h}]=\langle\ee^{\int \dd t \;{\bf h}\cdot\beeta}\rangle=
\exp\left[\nu\int\dd t\left(\langle\ee^{{\bll}\cdot {\bf h}(t)}\rangle_p-1\right) \right] ,
\end{equation}
where $\langle\cdot\rangle_p$ denotes an average with respect to $p({\bll})$ only. A key property is that the $n$-th order cumulant of $\beeta$ is nonzero only when the arguments are at equal times:
\begin{equation}\label{noisecumulants}
	\langle\eta^{\alpha_1}(t_1)\ldots\eta^{\alpha_n}(t_n)\rangle_c=\kappa^{(n)}_{\alpha_1,\ldots,\alpha_n}\delta(t_1-t_2)\ldots\delta(t_{n-1}-t_n) ,
\end{equation}
where $\kappa^{(n)}_{\alpha_1,\ldots,\alpha_n}=\nu\langle
\ell^{\alpha_1}\ldots \ell^{\alpha_n}\rangle_p$. The
$\alpha_j$'s denote arbitrary space directions: $\alpha_i= 1,\dots,d$ where $d$ is the number of spatial dimensions. In the following, we consider that only the even cumulants of the noise are non-zero. 

The celebrated Gaussian white noise is recovered in the scaling limit
$\nu\to\infty$ and $\langle \ell^\alpha \ell^\beta \rangle_p\to 0$,
with the effective diffusion constant $\gamma^{-1} T= \nu\langle {\bll
}^2\rangle_p/2d$ being fixed. In the latter scaling limit, the dynamics is equilibrium and, for
instance, the fluctuation-dissipation theorem follows. For the purpose
of comparison to equilibrium, we will stick to the notation
$T=\gamma(2d)^{-1}\nu \langle {\bll}^2\rangle_p$ even if out of
equilibrium. One must keep in mind that $T$ then loses its
thermodynamic meaning of a temperature. However, a curiosity of
Eq.~\eqref{LangevinNGunder} is that it nevertheless preserves some
sort of an ``equipartition'' theorem, according to which $\langle
m\bv^2/2\rangle=dT/2$ and $\langle\br\cdot\p_\br U\rangle=dT$.

Just as its Gaussian counterpart, Eq.~\eqref{LangevinNGunder} can also be considered in the overdamped limit. In that limit, somewhat unphysical features emerge that only $m\neq 0$ helps regularize. Considering the overdamped version,
\begin{equation}\label{LangevinNGover}
	\gamma \bv={\bf F}+\gamma\beeta ,
\end{equation}
one can see that after a given pulse the dynamics is the deterministic
gradient descent. Then, after a typical time $\nu^{-1}$, an
instantaneous pulse occurs again, with an infinite amplitude. For
finite forces $\bf F$, the infinite amplitude will always win over,
and this leads to the particle jumping instantaneously from one place
to another, possibly flying over existing obstacles. In practice, this
makes simulations particularly difficult in that limit. In the
Gaussian limit, such events become rarer due to the vanishing of the
hopping amplitude, but it is a well-known fact that the Brownian
trajectory is non differentiable and that sampling a Gaussian white
noise too can induce, however rarely, unphysical displacements. We
will get back to these caveats when considering interacting particles
in Sec.~\ref{sec:collective}.

Using the Kramers-Moyal expansion, the master equation for the probability $P(\bx,t)$ that the particle lies at $\br(t)=\bx$ reads
\begin{equation}\label{mastereq}
\p_t P=\gamma^{-1}\p_\bx\cdot ({\bf F}P)+\p_{\alpha_1}D_{\alpha_1}P ,
\end{equation}
where we have used the notation 
\begin{equation}\label{eq:DD}
	D_{\alpha_1}=\sum_{n\geq 2}\frac{(-1)^n\kappa^{(n)}_{\alpha_1,\ldots,\alpha_n}}{n!}\frac{\p^{n-1}}{\p x_{\alpha_2}\ldots\p x_{\alpha_n}} ,
\end{equation}
that generalizes the diffusive gradient $ \gamma^{-1}T\p_\bx$~\cite{vankampen,risken,gardiner}. A somewhat more formal way of denoting this operator is
\begin{equation}\label{eq:L}
	{\bf D}_\bx=\nu\frac{\tilde{p}(i\p_\bx)-1}{i\p_\bx} ,
\end{equation}
where $\tilde{p}(\bk)=\int_{\boldsymbol\ell} \ee^{-i\bk\cdot{\bll}}p({\bll})$. The last term in~\eqref{mastereq} also stems from the more intuitive master equation balance
\begin{equation}
	\p_{\alpha_1}D_{\alpha_1}P=\nu\int_{\bll}p(\bll)\left[P(\bx-\bll,t)-P(\bx,t)\right] .
\end{equation}
In what follows, we consider that $\beeta$ is a symmetric process where the jump distribution is isotropic.

% -------------------------------------------------------------------------------

\subsection{Specific jump distributions}\label{examplesofdistributions}

For the sake of clarity, we consider a specific jump distribution given by
\begin{equation}\label{eq:Gamma}
p ({\bll}) = {\mathcal N}_{d\alpha} (\ell / a )^{\alpha-1} \ee^{ - \ell / a } ,
\end{equation}
where $d+\alpha>1$, and $\ell=|{\bll}|$. The normalization constant reads
\begin{equation}
{\mathcal N}_{d\alpha} = [ \Omega_d a^d \Gamma ( \alpha + d - 1 ) ]^{-1} .
\end{equation}
The solid angle in $d$ dimension $\Omega_d = 2\pi^{d/2}/\Gamma(d/2)$ is written here in terms of the Euler Gamma function $\Gamma$. Such distributions typically emerge when considering that the jump length is the sum of $\alpha$ Poisson processes. We recover an exponential distribution for $\alpha = 1$, and the distribution gets all the more peaked around its average value as $\alpha$ increases. For such a distribution $\langle{\bll}^2\rangle_p=a^2(d+\alpha)(d+\alpha-1)$, so that the diffusion constant $D= \gamma^{-1}T=\nu\langle {\bll}^2\rangle_p/(2d)$ is of order $\nu a^2$.

To obtain the corresponding form of the operator ${\bf D}_\bx$ in Eq.~\eqref{eq:L}, we use the following expressions of the spatial Fourier transforms for the isotropic distributions ~\eqref{eq:Gamma}:
\begin{equation}
	\begin{aligned}
		d=1: \quad \tilde{p} (q) &= \int_{-\infty}^\infty \ee^{i q \ell} p (\ell) \dd \ell
		\\
		&= \f{ ( 1 - \ii a q )^\alpha + ( 1 + \ii a q )^\alpha }{2 \brt{ 1 + ( a q )^2 }^\alpha } ,
		\\
		d=2: \quad \tilde{p} (q) &= 2 \pi \int_0^\infty \ell J_0 ( q \ell ) p(\ell) \dd \ell
		\\
		&= {_2}F_1 \brt{ \f{1+\alpha}{2} , \f{2+\alpha}{2} ; 1 ; - (aq)^2 } ,
		\\
		d=3: \quad \tilde{p} (q) &= 4 \pi \int_0^\infty \f{\ell}{q} \sin(q\ell)p(\ell) \dd \ell
		\\
		&= \f{ \sin \brt{ (1+\alpha) \arctan(aq) } }{ a q (1+\alpha) \brt{ 1 + (aq)^2 }^{\f{1+\alpha}{2}} } ,
	\end{aligned}
\end{equation}
where $J_0$ and ${_2}F_1$ respectively denote the Bessel function of the first kind and the Gauss hypergeometric function. We report in Tab.~\ref{tab:L} the explicit form of ${\bf D}_\bx$ for some specific values of $d$ and $\alpha$.

\begin{table}
\caption{\label{tab:L}
Explicit form of the operator $\p_\bx\cdot{\bf D}_\bx/\nu a^2$ in~\eqref{eq:L}, expressed in terms of $L_0 = 1 - a^2 \p_\bx^2 $ for the jump distribution~\eqref{eq:Gamma}. We consider some specific values of the spatial dimension $d$ and of the jump parameter $\alpha$.
}
\centering
\begin{ruledtabular}
\begin{tabular}{cccc}
\\
$\df{\p_\bx\cdot{\bf D}_\bx}{\nu a^2}$ & $d = 1$ & $d=2$ & $d=3$
\\\\
\colrule
\\
$\alpha = 0$ & --- & $ \dfrac{L_0^{-1/2} - 1}{a^2}$ & $ \dfrac{\p_\bx^2}{L_0} $
\\\\
$\alpha=1$ & $ \dfrac{\p_\bx^2}{L_0} $ & $ \dfrac{1}{a^2} \left( L_0^{-3/2} - 1 \right)$ & $\dfrac{L_0 + 1}{L_0^2} \p_\bx^2 $
\\\\
$\alpha=2$ & $ \dfrac{2 + L_0}{L_0^2} \p_\bx^2 $ & $ \dfrac{1}{a^2} \left( \dfrac{ 3 - L_0 }{2 L_0^{5/2}} -1 \right) $ & $\dfrac{3 L_0 (L_0 + 1)+4}{3L_0^3} \p^2_\bx $
\\\\
\end{tabular}
\end{ruledtabular}
\end{table}

% -------------------------------------------------------------------------------

\subsection{Harmonic trap}
It is a well-documented fact, as reviewed by Solon {\it et al.}~\cite{Solon2015}, that both for RTPs and ABPs evolving in a quadratic potential, there can be an overshoot
of the probability to find the particle at a finite distance from the
center of the trap. An active particle with a finite propulsive force
${\bf F_p}$ indeed has a horizon $r_{\rm h}=|{\bf F_p}|/k$ for a trapping force
${\bf F}_{\rm trap}({\bf r})=- k {\bf r}$. When the time taken by the
particle to cross the trap is much shorter than the persistence time
of the propulsive force, the particle spends most of its time at the
horizon. The density profile in such cases is not that of a simple
decay from a peak at the center of the potential well, but it is actually
peaked at ${\bf r}\simeq r_{\rm h}$. Interestingly, this is not 
observed in AOUPs where the stationary distribution remains a
Gaussian~\cite{szamel2014PRE} (an equilibrium one, at
that~\cite{PhysRevLett.117.038103}). Thus, a natural question  is whether non-Gaussian white noise alone is responsible for a
non-monotonous density profile at odds with the intuition gained from
equilibrium. The answer is no, but there are some shared features. For
non-Gaussian but white noise, similar calculations have been done in
the past, but these do not really apply to the modeling of active
particles. For the example worked out in~\cite{dubkov2005generalized}
a L\'evy-type distribution with exponent $\alpha$ is obtained for the
position probability distribution function (pdf) of a particle in one
space dimension. This holds for a non-Gaussian white noise that is a
symmetric $\alpha$-stable L\'evy process, which is rather far from the
sort of non-Gaussian noise that is relevant to active particles. In
the latter, a typical hopping scale $a$ exists, as for instance in the
jump distributions discussed in
Sec.~\ref{examplesofdistributions}. For such jump distributions, it is
actually possible to find the Fourier transform of the position pdf of
a particle in a harmonic well $V(\br)=k \br^2 / 2$ with stiffness
$k$. The results are summarized in Tab.~\ref{tab:Pss}. They show the
Fourier transform of the steady state distribution $P_\text{ss}(\br)$,
defined by
\begin{equation}\label{defPss}
	\tilde P_\text{ss} ( \bq ) = \underset{t\to\infty}{\lim} \langle \ee^{ i \bq \cdot \br (t) } \rangle ,
\end{equation}
Introducing the response function $\chi(t)=\Theta(t)\ee^{-t/\tau_\R}$, with $\tau_\R= \gamma/ k$, we may rewrite the Langevin equation~\eqref{LangevinNGover} for ${\bf F}=-k\br$ as
\begin{equation}
	\br(t)=\int_{-\infty}^t\chi(t-u)\beeta(u)\dd u  .
\end{equation}
Therefore, using a cumulant expansion in Eq.~\eqref{defPss}, we arrive at
\begin{equation}\label{cumulantexpansionPss}
	\begin{aligned}
		\ln {\tilde P}_\text{ss} ( \bq ) &= \underset{t\to\infty}{\lim} \sum_{n=1}^\infty \frac{i^n}{n!} q_{\alpha_1} \cdots q_{\alpha_n}
		\\
		& \quad \times \int_{-\infty}^t \langle{ \eta_{\alpha_1} (t_1) \cdots \eta_{\alpha_n} (t_n) } \rangle_c
		\\
		& \quad \times \chi ( t - t_1 ) \cdots \chi ( t - t_n ) \dd t_1 \cdots \dd t_n ,
	\end{aligned}
\end{equation}
where the $\alpha_i$ indices are summed over and run from $1$ to $d$. Substituting the expression for the noise cumulants~\eqref{noisecumulants} into Eq.~\eqref{cumulantexpansionPss}, we obtain
\begin{equation}
\ln {\tilde P}_\text{ss} ( \bq ) = \nu \int_0^\infty\dd u \left( \tilde{p} ( \bq \chi(u) ) - 1\right) .
\end{equation}
Finally, using the change of variable $u\to s=\chi(u)$ the steady profile can be expressed as
\begin{equation}\label{eq:Pss}
\ln \tilde P_\text{ss}( \bq ) = \nu\tau_\R \int_0^1 \frac{\dd s}{s} \left( \tilde{p}(\bq s) - 1 \right) .
\end{equation}
One could directly check that Eq.~\eqref{eq:Pss} is indeed a solution
of Eq.~\eqref{mastereq} for ${\bf F}=-k\br$. We report in
Tab.~\ref{tab:Pss} explicit analytic expression for some specific
values of $\alpha$ and $d$. In $d=2$, the steady-state distribution
can always be explicitly computed as:
\begin{equation}
	\begin{aligned}
		\ln \tilde P_\text{ss} (\bq) &= - \f{\nu\tau_\R(1+\alpha)(2+\alpha)(aq)^2}{8}
		\\
		&\quad \times {_4}F_3 \brt{ 1, 1, \f{3+\alpha}{2}, \f{4+\alpha}{2} ; 2, 2, 2; -(aq)^2 } .
	\end{aligned}
\end{equation}
\begin{table}
\caption{\label{tab:Pss}
Space Fourier transform of the steady density profile in Eq.~\eqref{eq:Pss} for a $\alpha$-Gamma jump distribution in Eq.~\eqref{eq:Gamma}. We take specific values of the spatial dimension $d$ and the parameter $\alpha$.
}
\centering
\begin{ruledtabular}
\begin{tabular}{ccc}
${\tilde P}_\text{ss}$ & $d = 1$ & $d=3$
\\
\colrule
\\
$\alpha = 0$ & --- & $ \df{1}{\left[ 1 + (q a)^2 \right]^{ \f{\nu\tau_\R}{2} } }$
\\\\
$\alpha=1$ & $ \df{1}{\left[ 1 + (q a)^2 \right]^{ \f{\nu\tau_\R}{2}}} $ & $ \df{\ee^{ - \f{\nu\tau_\R}{2} \f{ (q a)^2 }{ 1 + (q a)^2 } }}{\left[ 1 + (q a)^2 \right]^{ \f{\nu\tau_\R}{2} }}  $
\\\\
$\alpha=2$ & $\df{\ee^{ - {\nu\tau_\R} \f{ (q a)^2 }{ 1 + (q a)^2 } } }{\left[1 + (q a)^2 \right]^{ \f{\nu\tau_\R}{2}}} $ & $  \df{\ee^{ - \frac{\nu\tau_\R}6 \f{ (q a)^2 \left[ 7 + 5 (q a)^2 \right]}{ [1 + (q a)^2]^2 } }}{\left[1 + (q a)^2 \right]^{ \f{\nu\tau_\R}{2}}} $
\\\\
\end{tabular}
\end{ruledtabular}
\end{table}
The corresponding distributions in real space show interesting
features like exponential tails (see~\cite{fodor_active_2016} for a
more mathematical discussion). The discussion depends on the value of
the $\nu\tau_\R$ combination, whose physical meaning is clear: the
higher this number the more frequent the white noise has stricken
during the typical relaxation times within the well.

When the relaxation within the well does not have time to proceed, for
$\nu\tau_\R\lesssim 1$, the center of the well becomes underpopulated
with respect to a Gaussian, and the distribution actually becomes
convex at the origin where it develops a cusp along with fat
tails. For instance, for $\alpha=1$, in one space dimension, an
explicit form of $P_\text{ss}$ can be found
\begin{equation}
	P_\text{ss}(x) \sim \Big(\frac{|x|}a\Big)^{(\nu\tau_\R-1)/2}K_{(\nu\tau_\R-1)/2}\Big(\frac{|x|}a\Big) ,
\end{equation}
where $K_n$ is the modified Bessel function of the second kind.  A
cusp does develop at the origin when $\nu\tau_\R<2$. This regime
mirrors that in which RTPs or ABPs exhibit a probability overshoot
away from the center of the well small $\nu$ means large persistence
time. A non-Gaussian white noise alone, however, is not sufficient to
produce an overshoot of the position pdf a finite distance away from
the bottom of the harmonic well for the choices of $p(\ell)$ that we
have tested. This is probably due to the absence of any mechanism to
select a specific length scale in our non-Gaussian models, as opposed
to ABPs and RTPs where the depletion of the center of the well leads
to an accumulation at the horizon $r_{\rm h}$. Here, the depletion instead
leads to fat tails.

In the opposite regime of small $a$ at fixed $\nu \tau_r a^2$, and
hence $\nu \tau_r \gg 1$, one recovers the Gaussian behavior $\tilde
P_\text{ss}(\bq) = \ee^{ - \frac{T q^2}{2k}}$. This regime mirrors
ABPs and RTPs which also behave as equilibrium particles in the limit
of vanishing persistence~\cite{Solon2015}. Finally, note that,
irrespective of the specific jump distribution $p(\bll)$,
equipartition holds in the sense that $k \langle \br^2\rangle/2=dT/2$
in $d$ space dimensions.

\begin{figure}
\centering
\includegraphics[width=\columnwidth]{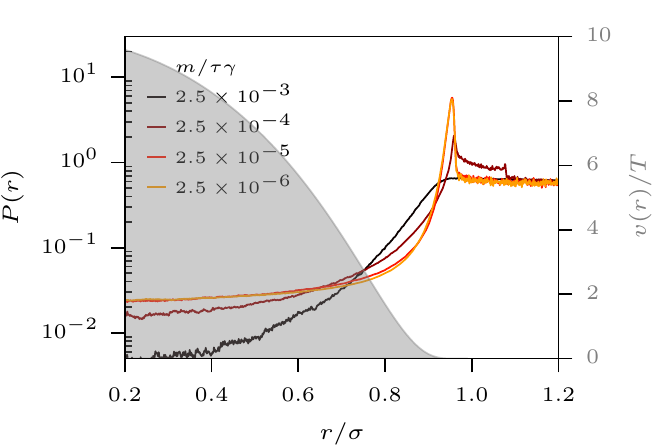}
\caption{\label{mass}
	Numerical results of the distribution of position for a particle subjected to a non-Gaussian white noise in two dimensions. The shaded region represents the form of the potential. The noise has a jump distribution $ p({\boldsymbol\ell}) \sim \ee^{ - \ell / a } $, and the particle evolves in the potential $v(r)=\eps \exp\brt{- 1/ ((\sigma/r)^2 -1) }\Theta(\sigma-r)$. The position pdf is reported as a function of the distance from the center of the potential for different values of $m/\tau\gamma$, where $\tau=\gamma\sigma^2/\varepsilon$: it converges at small values of $m/\tau\gamma$. Parameters: $T=1$, $\gamma=1$, $\eps=10$, $\sigma=1$, $\nu=20$.
}
\end{figure}

\subsection{Accumulation at boundaries}

With a view to gaining further intuition on the effect of a
non-Gaussian white noise, we continue our exploratory investigations
by considering an independent particle interacting with a fixed
obstacle. The obstacle is modelled by an external repulsive potential with range $\sigma$ and energy scale $\eps$, of the form $U({\bf r}) = v(|{\bf r}|) \Theta(\sigma-|\br|)$. The potential $v$ is either harmonic $v(r)=\varepsilon (1-r/\sigma)^2$ or exponential $v(r)=\varepsilon \exp\brt{- 1/ ((\sigma/r)^2-1) }$. We restore a non-zero mass for numerical purposes, as discussed in Sec.~\ref{sec:lang}. To probe the overdamped regime, we focus on small values of the inertial time $m/\gamma$ compared with the obstacle relaxation time $\tau=\gamma\sigma^2/\varepsilon$: the distribution indeed converges to a fixed profile, as shown in Fig.~\ref{mass}.

\begin{figure*}
\centering
\includegraphics[width=\columnwidth]{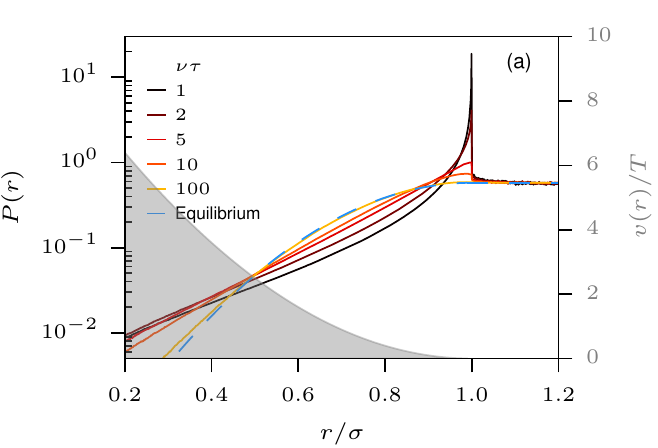}
\hfill
\includegraphics[width=\columnwidth]{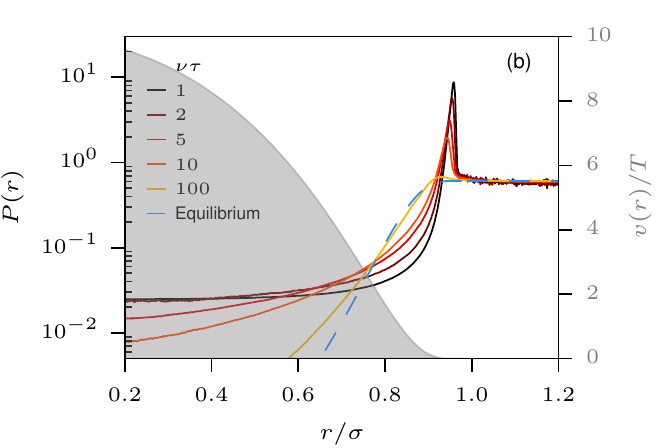}
\vskip.2cm
\includegraphics[width=\columnwidth]{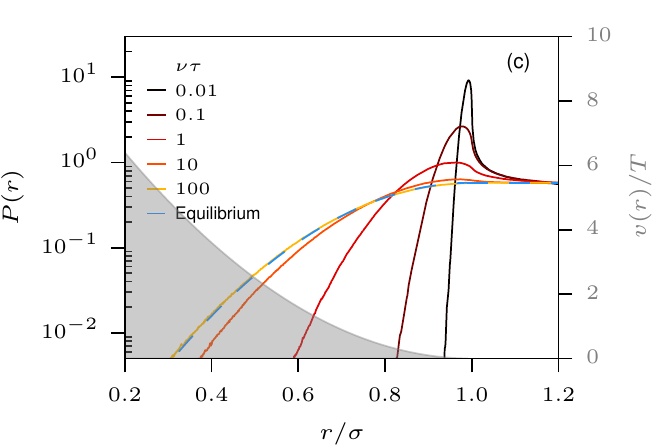}
\hfill
\includegraphics[width=\columnwidth]{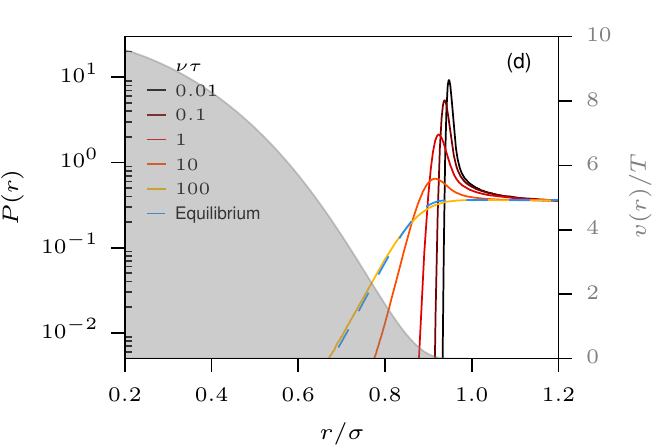}
\vskip.2cm
\includegraphics[width=\columnwidth]{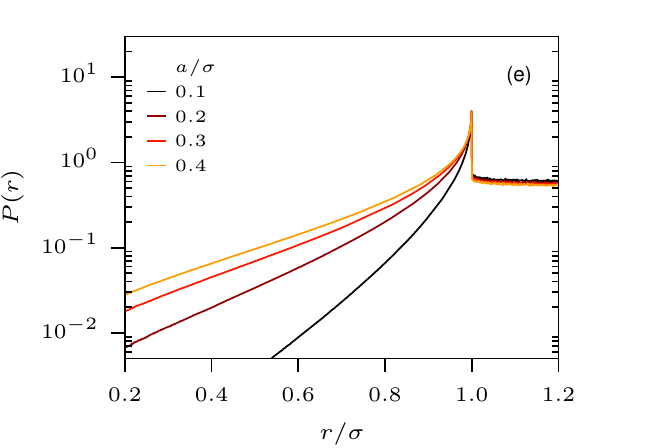}
\hfill
\includegraphics[width=\columnwidth]{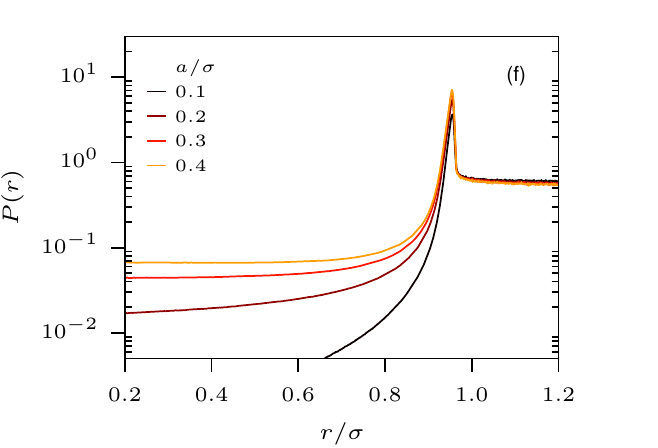}
\caption{\label{obstacle}
	Radial distribution of particle position away from the obstacle center. The shaded region represents the obstacle potential: harmonic $v(r)=\eps(1-r/\sigma)^2$ in (a, c, e) and exponential $v(r)=\eps \exp\brt{- 1/ ((\sigma/r)^2-1) }$ in (b, d, f). The blue dashed line in (a-d) refers to the equilibrium limit. The remarkable feature is the accumulation
of particles in the vicinity of the obstacle without the need to invoke memory effects, at a location where the equilibrium profile.
	(a--b)~Non-Gaussian white noise for different values of $\nu\tau$, where $\tau = \gamma \sigma^2 / \eps$.
	Parameters: $T=1$, $\gamma=1$, $\eps=10$, $\sigma=1$, $m=10^{-4}$.
	(c--d)~Exponentially correlated Gaussian noise with persistence time $1/\nu$ (AOUPs). Same parameters as in~(a--b).
	(e--f)~Non-Gaussian white noise for different values of $a/\sigma$ at fixed $ \nu \tau = 2 $. Since the temperature $T\sim\nu a^2$ varies, the potential $v(r)/T$ is not drawn here. All other parameters are the same as in~(a--b).
}
\end{figure*}

We measure the evolution of the radial distribution of the particle position away from the obstacle center. The obstacle is located at the center of a 2D box with periodic boundary conditions. For different values of $\nu\tau$, we compare the profile for non-Gaussian white noise, with jump distribution $ p({\boldsymbol\ell}) \sim \ee^{ - \ell / a } $, and the one for AOUPs with persistence time $1/\nu$. The accumulation at the obstacle boundary $r=\sigma$, shown in Figs.~\ref{obstacle}(a-b) for a non-Gaussian white noise, is qualitatively analog to that found with a persistent noise, reported in Figs.~\ref{obstacle}(c-d): it is more and more peaked as $\nu\tau$ decreases. This is consistent with previous results for persistent active particles~\cite{Elgeti2009,Tailleur2009,Maggi:15,Das2018}. For a harmonic obstacle, the distribution is singluar at $r=\sigma$ for non-Gaussian noise, at variance with the persistent case, and a cusp appears for $\nu\tau<5$, reminiscent of the profile under harmonic confinement. When increasing $a/\sigma$ at fixed $\nu\tau$, which amounts to increasing the temperature $T\sim\nu a^2$ as shown in in Figs.~\ref{obstacle}(e-f), the particle probes deeper regions of the potential, as expected.

Overall, our results support that the particle is effectively attracted to
the obstacle boundary for small $\nu\tau$. In this regime, the particle has ample time to go down the potential wall by steepest descent between two successive ``pulses'', with
typical time $\tau=\gamma\sigma^2/\varepsilon$. Hence it effectively accumulates down the potential instead of exploring the whole available space uniformly. In short, if the particle ever goes up the wall, it immediately goes down, hence the
probability increases right at the edge of the obstacle.  Assuming the obstacle can be viewed as a fixed particle, this suggests that effective two-body attraction could emerge in an assembly of particles driven by a non-Gaussian white noise, even though bare interactions are repulsive. A related question is whether these attractive effects, if present, are sufficiently strong to induce MIPS.

% ===============================================================================

\section{Collective dynamics}\label{sec:collective}

\subsection{Quorum sensing interactions}

To address interactions between particles, we first consider the case where
the statistics of self-propulsion depend on the local
density. Such quorum-sensing interactions are relevant to model
cells that adapt their motility to their local
environments~\cite{Miller2001}, leading to rich collective
behaviors~\cite{Liu2011}. For persistent self-propelled particles, a
propulsion speed decreasing as the local density increases has been
shown to lead to MIPS~\cite{PhysRevLett.100.218103, Tailleur:13, Bialke2013}. Furthermore, quorum-sensing interactions can be seen as an effective description of the kinetic slowing down induced by repulsive pairwise forces between
particles~\cite{Thompson2011, Bialke2013, Tailleur:13, soto2014, Slowman2016},
despite some important qualitative differences between these
models~\cite{solon2015pressure, solon2018generalized}.

In the context of a non-Gaussian white noise, we model the dynamics
of an individual particle by
\begin{equation}
	\gamma \frac{\dd \bx_i}{\dd t} = {\beeta}_i ,
\end{equation}
with the important difference that, now, the noise cumulants $\kappa_{\alpha_1\dots\alpha_n}^{(n)}$ in Eq.~\eqref{noisecumulants} depend on the local density $\rho({\bf x}, t) = \sum_i \delta^{(d)}(\bx-\bx_i(t))$:
\begin{equation}
	\kappa_{\alpha_1\dots\alpha_n}^{(n)} (\rho) = \nu \int \ell^{\alpha_1} \cdots \ell^{\alpha_n} p_\rho (\boldsymbol \ell) \dd {\boldsymbol \ell} .
\end{equation}
In practice, this is implemented by assuming that the jump distribution $p_\rho(\bll)$ itself is affected by $\rho$. Besides, each individual noise $\beeta_i$ remains independent between particles. The corresponding dynamics for the average density $\hat\rho({\bf x}, t) = \langle\rho({\bf x}, t)\rangle$ reads
\begin{equation}
	\p_t \hat\rho = \sum_{n\geq1} \f{ (-1)^n }{n!} \p_{x_{\alpha_1}} \cdots \p_{x_{\alpha_n}} \big[ \hat\rho \kappa_{\alpha_1\dots\alpha_n}^{(n)} (\hat\rho) \big] ,
\end{equation}
which can also be written as
\begin{equation}
	\p_t \hat\rho = \nu \int \sum_{n\geq1} \f{(- {\boldsymbol\ell} \cdot \p_{\bf x})^n}{n!} \brt{ p_\rho({\boldsymbol\ell}) \hat\rho } \dd {\boldsymbol\ell} .
\end{equation}
The emergence of a motility-induced phase separation at large scale can then determined from a linear stability analysis, in the spirit of~\cite{Tailleur:13, Solon2015}.

To do so, we consider fluctuations around the homogeneous profile $\rho_0$ and work to linear order in $ \delta \rho = \hat\rho - \rho_0 $:
\begin{equation}
		\p_t \delta \rho = \nu \int \dd {\boldsymbol\ell} \left.\brt{ p_\rho ({\boldsymbol\ell}) + \rho_0 p_\rho' ({\boldsymbol\ell}) }\right|_{\rho_0} \sum_{n\geq1} \f{(- {\boldsymbol\ell} \cdot \p_{\bf x})^n}{n!} \delta \rho ,
\end{equation}
where $ p_\rho' = \dd p_\rho / \dd\rho $. We infer the dynamics of the Fourier modes $ \delta \rho_{\bf k} (t) = \int \delta \rho ({\bf x}, t) \ee^{\ii {\bf k} \cdot {\bf x} } \dd {\bf x} $ as
\begin{equation}
	\begin{aligned}
		\p_t \delta \rho_{\bf k} &= \nu \delta \rho_{\bf k} \int \dd {\boldsymbol\ell} \left.\brt{ p_\rho ({\boldsymbol\ell}) + \rho_0 p_\rho' ({\boldsymbol\ell}) }\right|_{\rho_0} \sum_{n\geq1} \f{( \ii {\boldsymbol\ell} \cdot {\bf k})^n}{n!}
		\\
	  &= \nu \delta \rho_{\bf k} \int \dd {\boldsymbol\ell} \left.\brt{ p_\rho ({\boldsymbol\ell}) + \rho_0 p_\rho' ({\boldsymbol\ell}) }\right|_{\rho_0} \pnt{ \ee^{ \ii {\boldsymbol\ell} \cdot {\bf k} } - 1 }
		\\
	  &= \nu \delta \rho_{\bf k} \brt{ \pnt{ 1 + \rho_0 \f{\dd}{\dd\rho} } \left.\avg{ \ee^{ \ii {\boldsymbol\ell} \cdot {\bf k} } }\right|_{\rho_0} - 1 } ,
	\end{aligned}
\end{equation}
Assuming that the jump distribution does not have any angular dependence, so that only even moments of $\bll$ are nonzero, we deduce the following criterion for the occurrence long wavelength instabilities:
\begin{equation}
	\pnt{ 1 + \rho_0 \f{\dd}{\dd\rho} } \avg{ {\boldsymbol\ell}^2 } <  0 .
\end{equation}
This instability criterion does not depend on the jump rate $\nu$, as expected in the absence of any other time to compare it to.

\begin{figure}
\centering
\includegraphics[width=.6\columnwidth]{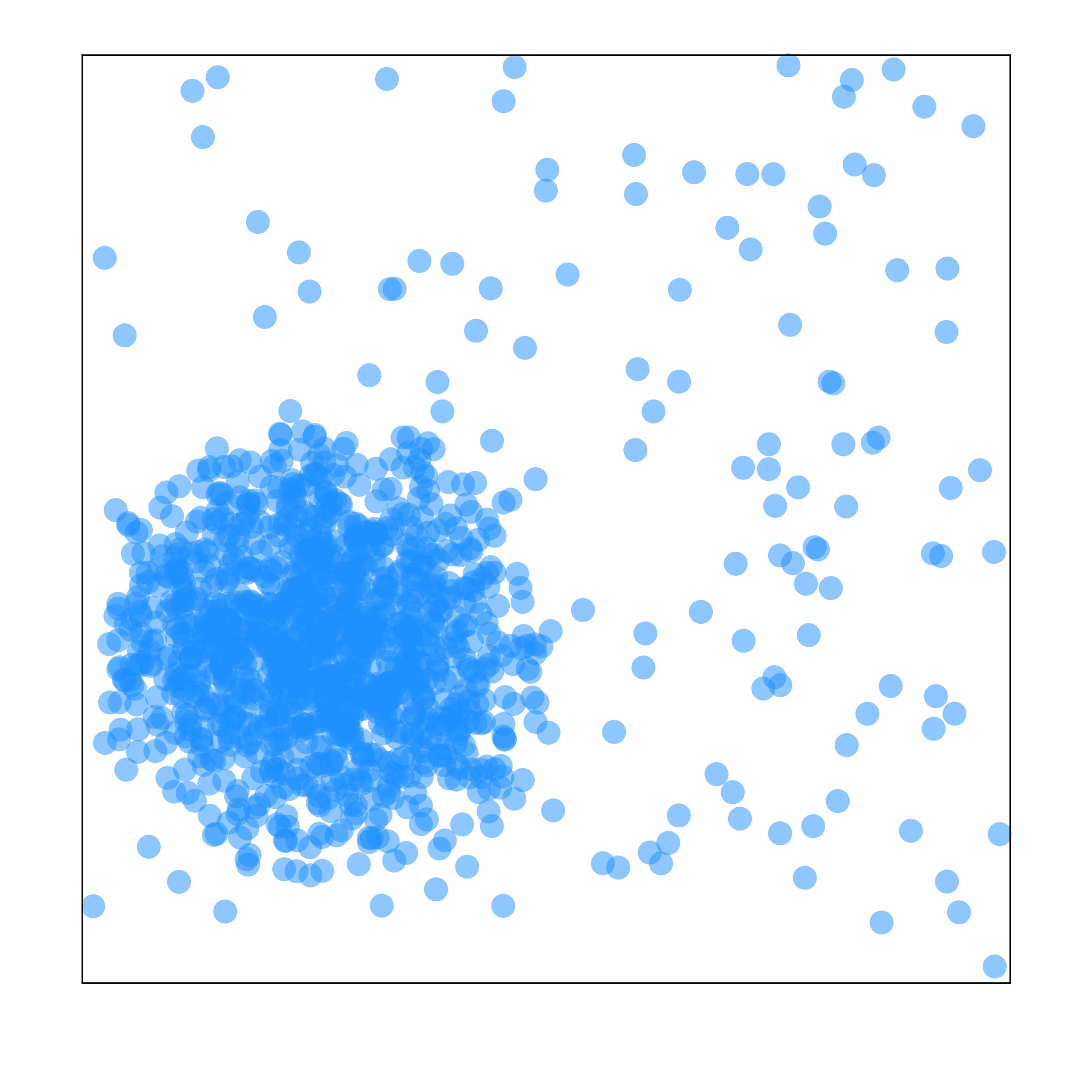}
\caption{\label{fig:QS}
	Phase separation for density-dependent jumping length $a(\rho)$, as given in~\eqref{eq:QS}, in a 2D box of size $L$, with periodic boundary conditions. For every particle, the blue circle denotes the area over which the local density is computed.
	Parameters: $\rho_0=0.6$, $\rho_\text{m}=0.6$, $\Delta\rho_0=0.006$, $L=40$, $a_\text{\tiny M}=10$, $a_\text{\tiny m}=1$.
}
\end{figure}

To assess the existence of phase separation, we perform simulations in a finite 2D box with periodic boundaries conditions. For simplicity, we choose the jump length to be fixed: $p_\rho({\boldsymbol\ell}) = \delta(|{\boldsymbol\ell}|-a(\rho))$, where $a$ depends on the local density as
\begin{equation}\label{eq:QS}
	a(\rho) = a_\text{\tiny M} + \frac{a_\text{m}-a_\text{\tiny M}}{2} \bigg[ 1 + \tanh \frac{\rho - \rho_\text{m}}{\Delta \rho} \bigg] .
\end{equation}
The typical values at low and high densities are $a_\text{\tiny M}$ and $a_\text{m}$, respectively. In practice, the local density is determined within a fixed radius
surrounding each particle. For appropriate values of
parameters, one indeed observe a complete phase separation between
dense and dilute regions, as reported in Fig.~\ref{fig:QS}.

% -------------------------------------------------------------------------------

\subsection{Pairwise forces and effective attraction}

To study the interplay between pairwise forces and non-Gaussian noises, we have performed 2D simulations of particles interacting {\it via} a two-body repulsive potential $U=\sum_{i\neq j} v(\br_i-\br_j)$, where $v(r) = \eps [(\sigma/r)^{12} - 2(\sigma/r)^6] \Theta(\sigma-r)$~\cite{WCA1971}. For similar interaction
potentials, RTPs, ABPs and AOUPs all display MIPS~\cite{Fily:12,Hagan:13,PhysRevE.91.042310,solon2018generalized}. Our
goal is to sort out the relative contributions of the persistence, on the one hand, and non-Gaussian statistics, on the other, which are typically intertwined in active particles.

We work at fixed particle density $\rho_0 = 0.6$, for which ABPs exhibit MIPS at large persistence, and we consider a fixed jump length $p({\boldsymbol\ell}) = \delta( |{\boldsymbol\ell}| - a)$. The equilibrium limit, corresponding to a Gaussian noise, is achieved as the hopping range $a\to 0$ and the hopping frequency $\nu\to\infty$ while keeping $T\sim\nu a^2$ fixed. Hence, we progressively drift away from equilibrium by slowing down the kicks at fixed temperature, namely by either reducing $\nu$ or increasing $a$ at fixed $\nu a^2$. Besides, to prevent particles from crossing each other when they should not, we use a finite yet small value of mass $m$. Note that this requires using extremely small time-steps, which significantly increases numerical cost.

\begin{figure}[b]
	\centering
	\includegraphics[width=\columnwidth]{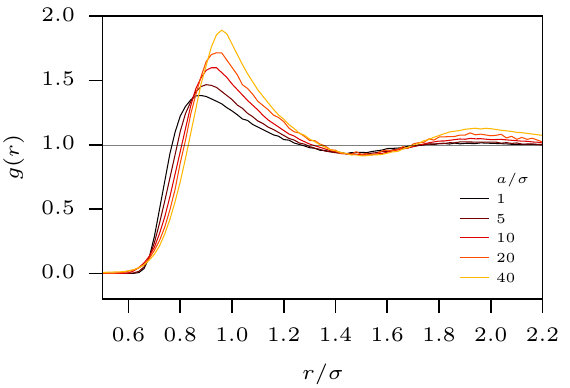}
	\caption{\label{fig:corr}
		Density pair correlation $g$ as a function of inter-particle distance $r$ scaled by particle diameter $\sigma$. The hopping range $a$ is varied at constant $\nu a^2 = 1$. Note the first and second peaks become all the more pronounced as we depart from equilibrium. Parameters: $\rho_0=0.6$, $m=10^{-3}$, $L=40$, $\gamma=1$, $\sigma=1$, $\varepsilon=10^2$.}
\end{figure}

The static structure is characterized by the two-body density correlation $g({\bf r}-{\bf r'}) = (1/\rho_0^2) \langle\rho({\bf r})\rho({\bf r}')\rangle$. We observe that the first two peaks of $g$, close to $r=\{\sigma, 2\sigma\}$, increase when departing away from the equilibrium regime, as shown in Fig.~\ref{fig:corr}. This suggests an increase of local order compatible with the emergence of motility-induced clustering. Note, however, that for the times and sizes accessible to our numerics, we could not observe a complete phase separation. We defer an extensive analysis of the corresponding finite size effects to future works.

% -------------------------------------------------------------------------------

\subsection{Generalized Dean-Kawasaki equation}

To describe collective effects, we now analyze the statistics of the fluctuating particle density $\rho$ for pairwise forces. The dynamics of $\rho$ can be obtained in the same vein as for a Gaussian white noise by using It\^o calculus~\cite{dean_langevin_1996, Kawasaki}. Yet, this derivation must be greatly revised due to the non-Gaussian nature of the noise. This has been considered two decades ago in the mathematical literature~\cite{DiPaola1992, DIPAOLA1993197, dipaolastochastic1993}. Appendix~\ref{beyonditostrato} gives the proper discretization scheme in physical terms, recently revived in \cite{PhysRevLett.108.210601}, which is consistent with differential calculus for a generic non-Gaussian noise.

The corresponding chain rule then leads to
\begin{equation}\label{continuity-1}\begin{split}
\p_t\rho=&-\p_{\bx}\cdot\sum_i\frac{\dd\br_i}{\dd t}*\delta(\bx-\br_i)\\
=&-\p_{\bx}\cdot\sum_i\left(-\gamma^{-1}\p_{\br_i}U+{\beeta}_i\right)*\delta(\bx-\br_i) ,
\end{split}\end{equation}
where the multiplicative noise signaled by the $*$ product, must be understood in terms of $\Delta\rho(\bx,t)=\rho(\bx,t+\Delta t)-\rho(\bx,t)$ and $\Delta{\beeta}_i=\int_t^{t+\Delta t}\dd\tau{\beeta}_i(\tau)$ as
\begin{equation}\begin{split}
\Delta\rho=&-\p_{\bx}\cdot\sum_i\left[-\gamma^{-1}\p_{\br_i}U\delta(\bx-\br_i)\right.\\&\left.+\frac{\ee^{\Delta{\beeta}_i\cdot\p_{\br_i}}-1}{\Delta{\beeta}_i\cdot\p_{\br_i}}\delta(\bx-\br_i)\Delta{\beeta}_i\right]\\
=&-\p_{\bx}\cdot\sum_i\left[-\gamma^{-1}\p_{\br_i}U\delta(\bx-\br_i)\right.\\&\left.+\frac{\ee^{-\Delta{\beeta}_i\cdot\p_{\bx}}-1}{-\Delta{\beeta}_i\cdot\p_{\bx}}\delta(\bx-\br_i)\Delta{\beeta}_i\right] .
\end{split}\end{equation}
This allows us to determine the Kramers-Moyal coefficients $\{K^{(k)}\}$, defined by
\begin{equation}
 K^{(k)}(\bx_1,\ldots,\bx_k)=\lim_{\Delta t\to 0}\frac{\langle\Delta\rho(\bx_1)\ldots\Delta\rho(\bx_k)\rangle}{\Delta t} .
\end{equation}
Once these coefficients are known, we can directly write an {\it It\^o-discretized} stochastic equation for $\rho$, in the following form
\begin{equation}\label{NGDK}
	\p_t\rho=K^{(1)}+\text{noise} .
\end{equation}
Here, the white noise is non-Gaussian, with zero-mean and cumulant amplitude given by $K^{(k)}$ for $k\geq 2$. A tedious but straightforward calculation leads to
\begin{equation}\begin{split}
K^{(1)}(\bx)=&\gamma^{-1}\p_\bx\cdot\left[\rho(\bx,t)\int_\by \p_\bx v(\bx-\by)\rho(\by,t)\right]\\
&-\p_\bx\cdot\sum_{p=0}^\infty\frac{(-1)^p\kappa^{(p+1)}}{(p+1)!}\p_\bx^p\rho ,
\end{split}\end{equation}
and, for $k\geq 2$,
\begin{equation}\label{cumul-NGDK}\begin{split}
	K^{(k)}&=(-1)^k\p_{\bx_1}\ldots\p_{\bx_k}\sum_{p_1,\ldots,p_k}\frac{(-1)^{p_1+\ldots+p_k}\kappa^{(p_1+\ldots+p_k+k)}}{(p_1+1)!\ldots (p_k+1)!}
	\\
	&\quad\times\p^{p_1}_{\bx_1}\ldots\p^{p_k}_{\bx_k}\rho\delta(\bx_1-\bx_2)\ldots\delta(\bx_{k-1}-\bx_k) .
\end{split}\end{equation}
Equations~(\ref{NGDK}-\ref{cumul-NGDK}) can be viewed as the non-Gaussian generalization of the Dean-Kawasaki equation. Again, we stress that, by construction, it is written in It\^o form. A somewhat more physically appealing form reads
\begin{equation}\label{eq:rho}
	\p_t \rho=-\p_\bx\cdot{\bf j} ,
\end{equation}
where the fluctuating current $\bf j$ is given by
\begin{equation}\label{eq:j}
	{\bf j}(\bx,t)=- {\bf D}_\bx\rho-\frac{1}{\gamma}\rho(\bx,t)\int_\by\p_\bx v(\bx-\by)\rho(\by,t)+\text{noise} .
\end{equation}
The notation ${\bf D}_\bx$, which generalizes the simple diffusive transport, has already been introduced in~\eqref{eq:DD}.

One can reformulate the Langevin equation~(\ref{NGDK}-\ref{cumul-NGDK}) in terms of a Martin-Siggia-Rose-Janssen-De Dominicis path integral. We demonstrate in Appendix~\ref{action} that the corresponding dynamical action can be written as
\begin{equation}\label{eq:action_def}
	\begin{aligned}
	S &= \int_{\bx, t} \brt{ \bar \rho \p_t \rho + \f{1}{\gamma} \p_\bx \bar \rho \cdot \int_\by \rho(\bx,t) \p_\bx v(\bx-\by) \rho (\by,t) }
	\\
	& \quad - \nu \int_{\bx, \bll, t} \rho(\bx,t) \brt{ \ee^{ \bar \rho (\bx+\bll,t) - \bar \rho (\bx,t) } - 1 } p(\bll) .
	\end{aligned}
\end{equation}
The It\^o discretization ensures that one does not have do deal with any Jacobian. We treat here $\rho$ as a well-behaved field of integration, though it is {\it a priori} a sum of delta functions centered around each particles. A formal proof that this is indeed legitimate for an ideal gas of Brownian particles has been given in~\cite{velenich_brownian_2008}. While~\eqref{eq:action_def} is fully general, it is also remarkably complex. As a consistency check, one can proceed directly from a Doi-Peliti approach using the second-quantized operators $a$ and $\bar a$~\cite{Doi1976, Peliti1985}, where the contribution for the particle hops reads
\begin{equation}
	\nu \int_{\bx, \bll} p(\bll) \brt{ \bar a (\bx+\bll,t) - \bar a (\bx,t) } a(\bx,t) .
\end{equation}
Using the density operators $\rho$ and $\bar \rho$ introduced by Grassberger as~\cite{grassberger_phase_1982}  $a=\ee^{-\bar\rho}\rho$ and $\bar a=\ee^{\bar\rho}$, one ends up with the same dynamic action~\eqref{eq:action_def}, as detailed in Appendix~\ref{action}.

% -------------------------------------------------------------------------------

\subsection{Density correlations: perturbative treatment}

The Langevin equation~(\ref{NGDK}-\ref{cumul-NGDK}), or alternatively the dynamic action~\eqref{eq:action_def}, provide a systematic toolbox to study $k$-point correlations $\rho^{(k)}$ defined as
\begin{equation}
	\rho^{(k)}(\bx_1,\ldots,\bx_k) = \Big\langle \sum_{i_1\neq\ldots \neq i_k}\delta(\bx_1-\br_{i_1})\dots\delta(\bx_k-\br_{i_k}) \Big\rangle
\end{equation}
through a BBGKY hierarchy~\cite{Hansen2013}.
For instance, the dynamics of the first non-trivial correlations can be written for $t'<t$ as
\begin{equation}\begin{split}
	\p_t \langle\rho(\bx',t')&\rho(\bx,t)\rangle=\p_\bx\cdot{\bf D}_\bx\langle\rho(\bx',t')\rho(\bx,t)\rangle\\
&+\gamma^{-1}\p_\bx\cdot\int_\by\p_\bx v(\bx-\by)\langle\rho(\bx',t')\rho(\bx,t)\rho(\by,t)\rangle .
\end{split}\end{equation}
In the limit $t'\to t^-\to\infty$, using
\begin{equation}\begin{split}
\langle&\rho(\bx)\rho(\bx')\rangle=\rho^{(2)}(\bx-\bx')+\rho_0\delta(\bx-\bx'), \\
\langle\rho(\bx')&\rho(\bx)\rho(\by)\rangle=\rho^{(3)}(\bx,\bx',\by)
+\delta(\bx-\bx')\rho^{(2)}(\bx,\by)\\&+\delta(\bx-\by)\rho^{(2)}(\bx,\bx')+\delta(\bx'-\by)\rho^{(2)}(\bx,\by)\\
&+\delta(\bx-\bx')\delta(\bx-\by)\delta(\bx'-\by)\rho^{(1)}(\bx) ,
\end{split}\end{equation}
we arrive at
\begin{equation}\label{BGY}\begin{split}
	0&=\p_\bx\cdot\big[{\bf D}_\bx\rho^{(2)}(\bx,\bx')+\gamma^{-1}\rho^{(2)}(\bx,\bx')\p_{\bx} v(\bx-\bx')\big]
	\\
	&\quad+\gamma^{-1}\p_\bx\cdot\int_\by\p_\bx v(\bx-\by)\rho^{(3)}(\bx,\bx',\by) .
\end{split}\end{equation}
This can be regarded as the non-Gaussian generalization of the BGY equation~\cite{Yvon1935,BornGreen1946}, originally introduced for Hamiltonian dynamics.

We now consider the low density regime where correlations of order $k>2$ are negligible and analytical progress is possible: the generalized BGY~\eqref{BGY} then reduces to an equation for pair correlations $g(r)=\rho^{(2)}(r)/\rho_0^2$ only. For a generic interaction potential $v(\br)$, no exact solution can be found. Using the Boltzmann-Gibbs weight as a reference
distribution, we expand the stationary state in powers of the non-Gaussianity of the applied noise.
A similar expansion was recently carried out within the framework of quantitative finance~\cite{popescu2015kramers} for a single degree of freedom.
Scaling position as ${\bf r}'={\bf r}/\sigma$, where $\sigma$ is a typical length scale such as the range of interactions, we get
\begin{equation}\label{BGY_g}
	0=\p_{\br'}\cdot(g \p_{\br'} v)+T\p_{\br'}^2 \bigg[1-c_{d,\alpha} \Big(\frac{a}{\sigma}\Big)^2\p_{\br'}^2 + {\cal O}\Big(\frac{a}{\sigma}\Big)^4\bigg] g ,
\end{equation}
where $c_{d,\alpha}$ is a dimensionless coefficient which depends on the spatial dimension $d$ and on the jump parameter $\alpha$. Using the Boltzmann weight as a reference distribution, we expand $g$ in powers of $a$ at fixed $T\sim\nu a^2$ as: $g(\br')\sim\ee^{-(v(\br')+ w(\br'))/T}$, where $w\to 0$ in the Gaussian limit. We deduce that $w={\mathcal O}(a/\sigma)^2$ and that it satisfies
\begin{equation}\label{twobodyeffectivepot}
	\frac{\p_{\br'} w}{c_{d,\alpha}} = \Big(\frac{a}{\sigma}\Big)^2 \bigg[\p_{\br'} - \frac{(\p_{\br'} v)}{T}\bigg]\bigg[\frac{(\p_{\br'} v)^2}{T}-\Delta_{\br'} v\bigg] .
\end{equation}
For a simple repulsive potential $v(\br)=\eps(\sigma/r)^{12}$, we find
\begin{equation}
	\begin{aligned}
		\frac{w(\br)}{c_{d,\alpha}} = 24\eps \pnt{\f{a}{\sigma}}^2 \bigg[& \frac{d-14}{2} \Big(\f{\sigma}{r}\Big)^{14} + \f{3(d-40)}{13} \f{\eps}{T} \Big(\frac{\sigma}{r}\Big)^{26}
		\\
		& - \f{36}{19} \Big(\f{\eps}{T}\Big)^2 \Big(\frac{\sigma}{r}\Big)^{38} \bigg] ,
	\end{aligned}
\end{equation}
where we have used $ \Delta_{\bf r} v = r^{1-d} \p_r \pnt{ r^{d-1} \p_r v } $. The leading contribution induces an attractive interaction at distances of the order of $\sigma$, with a strength proportional to $T/\nu$, similarly to AOUPs with persistence $1/\nu$~\cite{PhysRevLett.117.038103}. This suggests that effective attractive interactions are indeed to be expected in a dilute limit and for mildly non-Gaussian noise. This is consistent with our numerical observations in Fig.~\ref{fig:corr}.

% ===============================================================================

\section{Conclusion}

The dynamical evolution of an active particle is often modeled after a Langevin equation in which the noise source drives the particle away from equilibrium by displaying persistent properties along with non-Gaussian statistics~\cite{Marchetti:2013:RMP, Bechinger2016, Marchetti2018}. In this work, we have focused on the purely non-Gaussian effects of the active noise by discarding all memory effects. We have provided analytical results for the position distribution of a single particle in a harmonic trap. We have also reported some numerical evidence of accumulation at the boundaries of an obstacle. This supports that non-Gaussian effects alone can yield effective attraction from bare repulsion, similarly to the case of persistent particles~\cite{Martens2012, Solon2015, Maggi:15, Das2018}.

Considering particles interacting {\it via} quorum-sensing, we report the existence of a phase separation analog to MIPS~\cite{PhysRevLett.100.218103, Cates:15}. For pairwise forces, the effective attraction enhances spontaneous clustering, as testified by density correlations. Yet, within the explored range of parameters, we have not witnessed any phase separation for such interactions. To investigate collective effects, we have derived the stochastic density dynamics by extending standard procedures to non-Gaussian noise~\cite{dean_langevin_1996, Kawasaki}. In the dilute limit and for weakly non-Gaussian noise, a systematic expansion has confirmed the emergence of effective attraction from bare repulsive interactions, in line with our numerical results.

When driven by a non-Gaussian white noise, the dynamics in the presence of interactions operates far from equilibrium by breaking time reversal symmetry. It would be interesting to investigate deeper the consequences for a ratchet, where a current develops spontaneously in an asymmetric periodic potential, by analogy with persistent noises~\cite{Leonardo2011, Maggi2014b, Grosberg2017, Reichhardt2017}. Moreover, the properties of a heat engine subject to a non-Gaussian white noise could be explored, following~\cite{PhysRevLett.108.210601, Sagawa2014, Zakine2017}. In particular, the role of particle interactions in the performance of ratchets and engines, studied recently for driven and persistent particles~\cite{Dhar2015, Martin2018}, calls for deeper investigation in the non-Gaussian case.

% ===============================================================================

\acknowledgements

We warmly acknowledge discussions with Kiyoshi Kanazawa and Tomohiko Sano. While this work was completed FvW benefitted from a Yukawa Institute visiting professorship, and \'EF benefitted from the YITP financial support. We also acknowledge financial support from IRSES SoftActive. HH acknowledges financial support from by the Grant-in-Aid of MEXT for Scientific Research (Grant No. 16H04025). FvW is a member of the Institut Universitaire de France. JT acknowledges support from ANR grant Bactterns. \'EF benefits from an Oppenheimer Research Fellowship from the University of Cambridge, and a Junior Research Fellowship from St Catherine's College.

% ===============================================================================

\appendix

\section{Discretization issues}\label{beyonditostrato}

This Appendix is devoted to presenting, in a way familiar to the chemist or the physicist~\cite{vankampen,risken,gardiner} and based on the Kramers-Moyal expansion, the rules of stochastic calculus involving white but non-Gaussian noise. These rules are well-known to the mathematics community, which has its own language to express these~\cite{DiPaola1992,DIPAOLA1993197,dipaolastochastic1993} (see also \cite{Pirrotta2007127} for a more recent exposition). The present Appendix is also an alternative to the more recent presentation by Kanazawa {\it et al.}~\cite{PhysRevLett.108.210601}. To make things as pedagogical as possible, we begin with a general Langevin equation for a process $x(t)$ evolving according to
\begin{equation}\label{LangevinNG-formal}
\frac{\dd x}{\dd t}=A(x)+B(x)\eta(t)
\end{equation}
where the noise $\eta$ is characterized by its cumulants
\begin{equation}
\langle\eta(t_1)\ldots\eta(t_n)\rangle=\kappa^{(n)}\delta(t_1-t_2)\ldots\delta(t_{n-1}-(t_n)
\end{equation}
The experienced reader knows that, as such, Eq.~\eqref{LangevinNG-formal} needs to be supplemented by a discretization rule (without which it is meaningless) and the product $B\eta$ is best written using a warning sign $B*\eta$. By definition, the It\^o rule for understanding \eqref{LangevinNG-formal} reads
\begin{equation}
\Delta x=x(t+\Delta t)-x(t)=A(x(t))\Delta t+B(x(t))\int_t^{t+\Delta t}\dd\tau \eta(\tau)
\end{equation}
where $\Delta t$ is an infinitesimal discretization scale. In the sequel, we will often use the notation $\Delta\eta =\int_t^{t+\Delta t}\dd\tau \eta(\tau)$. It is easy to realize that
\begin{equation}
\langle(\Delta \eta)^k\rangle=\kappa^{(k)}\Delta t+o(\Delta t)
\end{equation}
and thus, as $\Delta t\to 0$
\begin{equation}
\frac{\langle \Delta x\rangle}{\Delta t}=A,\,\frac{\langle \Delta x^k\rangle}{\Delta t}=(B(x))^k\kappa^{(k)}\text{ for }k\geq 2
\end{equation}
and this explains the form of the Master or Fokker-Planck equation \eqref{mastereq} obtained for $B=1$ and $A=-\p_x  U$. Of course, depending on context, other discretization rules could be implemented on Eq.~\eqref{LangevinNG-formal}, and they would lead to different physical processes with different Fokker-Planck equations . For instance, the Stratonovitch rule would read
\begin{equation}\label{real-strato}\begin{split}
\Delta x=&x(t+\Delta t)-x(t)=A\Delta t+B\left(x(t)+\frac 12 \Delta x\right)\Delta\eta\\
=& A\Delta t+B\Delta\eta+\frac 12 B'\Delta x\Delta\eta+\ldots\\
=&A\Delta t+B\Delta\eta+\frac 12 B'B\Delta\eta^2+\ldots
\end{split}
\end{equation}
where the $\ldots$ stand for terms that are of lower order in the $\Delta t\to 0$ limit only when the noise is Gaussian.
For a Gaussian white noise, the Stratonovitch rule ensures that the chain rule is consistent with stochastic calculus. In other words, given an arbitrary function $f(x)$, with the Stratonovitch rule one may safely write that 
\begin{equation}\label{chain-rule-S}
\frac{\dd f}{\dd t}=f'(x(t))\frac{\dd x}{\dd t}=f'(A+B*\eta)=f'A+(f'B)*\eta
\end{equation}
where $(f'B)*\eta$ in the right-hand-side is to be understood in the Stratonovitch sense as long as $B*\eta$ in the evolution of $x(t)$ is as well. But \eqref{chain-rule-S} only holds for the Stratonovitch discretization and for $\eta$ a Gaussian white noise. This is not the case anymore for a generic non-Gaussian white noise. Another discretization rule plays this special role of being transparent to differential calculus. It is defined in the following way:
\begin{equation}\label{Strato-generalized}
\Delta x=x(t+\Delta t)-x(t)=A\Delta t+\frac{\ee^{B(x)\Delta\eta\frac{\dd}{\dd x}}-1}{B(x)\Delta\eta\frac{\dd}{\dd x}}B(x)\Delta\eta
\end{equation}
where the $t$ argument in $x(t)$ was omitted. Note that truncating the right-hand side in \eqref{Strato-generalized} to order $\Delta\eta^2$ leads one to recovering \eqref{real-strato}. To leading order in $\Delta t$, and for a Gaussian white noise, both discretization prescriptions are identical. Now, let's prove that the prescription Eq.~\eqref{Strato-generalized} is indeed transparent to differential calculus in the sense that differential calculus can blindly be used. We consider a function $f(x(t))$ and ask whether we have
\begin{equation}\label{transparence-calcul-diff}
\frac{\dd f}{\dd t}=f'(x(t))*\frac{\dd x}{\dd t}=f'*(A+B*\eta)=f'A+(f'B)*\eta
\end{equation}
Introducing a discretization scale $\Delta t$, we must evaluate $\Delta f=f(x+\Delta x)-f(x)$ in two ways. We introduce the generalized translation operator $\hat{T}_B$ defined by $\hat{T}_B[w]=\frac{\ee^{\Delta\eta B\frac{\dd}{\dd x}}-1}{\Delta\eta B\frac{\dd}{\dd x}}w(x)$. We begin with
\begin{equation}\label{chain-rule-1}\begin{split}
\Delta f=&f(x+\Delta x)-f(x)\\
=&f(x+A\Delta t+\hat{T}_B [B\Delta\eta])-f(x)
\end{split}\end{equation}
which we want to compare with the expression that would be deduced from the blind application of the chain rule:
\begin{equation}\label{chain-rule-2}
\begin{split}
\Delta f=&f'A \Delta t+\int_t^{t+\Delta t}(f'B)*\eta\\
=&f'A\Delta t+\hat{T}_B[f'B\Delta\eta]
\end{split}
\end{equation}
If we can prove that Eqs.~\eqref{chain-rule-1} and \eqref{chain-rule-2} are actually identical, then we will have established our result. First we note that for any function $f$,
\begin{equation}
\hat{T}_B[f'B\Delta \eta]=\left(\ee^{B\Delta\eta\frac{\dd }{\dd x}}-1\right)f
\end{equation}
which also means, for $f(x)=x$, that $\hat{T}_B[B\Delta\eta]=\left( \ee^{B\Delta\eta\frac{\dd }{\dd x}}-1 \right) x$, and we are left with the following identity between Eqs.~\eqref{chain-rule-1} and \eqref{chain-rule-2} to prove as $\Delta t\to 0$:
\begin{equation}
f\left(A\Delta t+\ee^{B\Delta\eta\frac{\dd}{\dd x}}x\right)=f' A\Delta t+\ee^{B\Delta\eta\frac{\dd}{\dd x}}f
\end{equation}
In the left-hand side, we first remark that
\begin{equation}\begin{split}
f\left(A\Delta t+\ee^{B\Delta\eta\frac{\dd}{\dd x}}x\right)=&f\left(\ee^{B\Delta\eta\frac{\dd}{\dd x}}x\right)\\&+f'\left(\ee^{B\Delta\eta\frac{\dd}{\dd x}}x\right)A\Delta t\\
&+ {\mathcal O}(\Delta t^2)\\
=&f\left(\ee^{B\Delta\eta\frac{\dd}{\dd x}}x\right)\\&+f'(x) A\Delta t+ {\mathcal O}(\Delta t^2)
\end{split}\end{equation}
We now prove the following exact identity (valid at arbitrary $\Delta t$):
\begin{equation}\label{jolie-identite}
f\left(\ee^{B\Delta\eta\frac{\dd}{\dd x}}x\right)=\ee^{B\Delta\eta\frac{\dd}{\dd x}}f(x)
\end{equation}
To do so we introduce the variable $y$ such that $\frac{\dd x}{\dd y}=B\Delta\eta$ and write $x=g(y)$ where we do not need the explicit form of $g$. In terms of the $y$ variable, Eq.~\eqref{jolie-identite} becomes
\begin{equation}\label{trans-op}
f\left(\ee^{\frac{\dd}{\dd y}}g(y)\right)=\ee^{\frac{\dd}{\dd y}}f(g(y))
\end{equation}
Using that for any function $h(y)$, $\ee^{\frac{\dd}{\dd y}}h(y)=h(y+1)$, which we apply to $h(y)=g(y)$ and to $h(y)=f(g(y))$, we have thus proved Eq.~\eqref{trans-op}, which in turn establishes Eq.~\eqref{jolie-identite} and thus ensures the equality between Eqs.~\eqref{chain-rule-1} and \eqref{chain-rule-2}. Hence, for a non-Gaussian white noise, the prescription rule Eq.~\eqref{Strato-generalized} allows for the blind use of differential calculus.

% -------------------------------------------------------------------------------

\section{Dynamic action}\label{action}

Our goal is to obtain an explicit expression of the dynamic action $\cA$ in terms of the jump distribution $\cW$ of the non-Gaussian noise. Following standard procedures, the action associated with the dynamics of the position density in~(\ref{eq:rho}--\ref{eq:j}) can be written as
\begin{equation}\label{eq:action}
	\begin{aligned}
		& \cA = \int_{\bx, t} \brt{ \bar \rho \p_t \rho + \f{1}{\gamma} \p_\bx \bar \rho \cdot \int_\by \rho(\bx,t) v(\bx-\by) \rho(\by,t) }
		\\
		& \, - \sum_{n=1}^\infty \f{1}{n!} \int_{ \bx^n, t } \bar \rho (\bx_1, t) \cdots \bar \rho (\bx_n, t) \avg{ \xi (\bx_1, t) \cdots \xi (\bx_n, t) }_c ,
	\end{aligned}
\end{equation}
where the noise term $\xi$ is written in terms of the microscopic noises $ \cur{ {\boldsymbol\eta}_i } $ and $\rho_i (\bx, t) = \delta \brt{ \bx - \bx_i(t) }$ as
\begin{equation}
	\xi (\bx, t) = - \sum_{i=1}^N \nabla \brt{ {\boldsymbol\eta}_i (t) \rho_i (\bx, t) } .
\end{equation}
The noise cumulants can be expressed in terms of the variation of density $\Delta \rho_j = \Delta \rho (\bx_j, t)$ during a time $\Delta t$ in the absence of potential as
\begin{equation}
	\avg{ \xi (\bx_1, t) \cdots \xi (\bx_n, t) }_c = \underset{ \Delta t \to 0 }{\lim} \f{ \avg{ \Delta \rho_1 \cdots \Delta \rho_2 } }{\Delta t} .
\end{equation}
The density variation in the absence of potential is given by
\begin{equation}\label{eq:drho}
\Delta \rho (\bx, t) = \sum_{i=1}^N \sum_{k=1}^\infty \f{ (-1)^k }{k!} \brt{ \Delta {\boldsymbol\eta}_i (t) \cdot \nabla }^k \rho_i (\bx, t) .
\end{equation}
The product of the density variations follows as
\begin{equation}
	\begin{aligned}
		\avg{ \Delta \rho_1 \cdots \Delta \rho_n } &= \nu \sum_{ k^n, \alpha^n, i^n } \avg{ \Delta \eta_{i_1 \alpha_1}^{k_1} \cdots \Delta \eta_{i_n \alpha_n}^{k_n} }
		\\
		& \quad \times \prod_{j=1}^n \f{ (-1)^{k_j} }{k_j!} \pnt{ \p_{i_j \alpha_j} }^{k_j} \rho_{i_j} (\bx_j, t) ,
	\end{aligned}
\end{equation}
where the elements in $\alpha^n = \cur{ \alpha_1, \dots, \alpha_n }$ run from $1$ to $d$, the ones in $k^n = \cur{k_1, \dots, k_n}$ from $1$ to $\infty$, and the ones in $i^n = \cur{i_1, \dots, i_n}$ from $1$ to $N$. Substituting the expression for the noise cumulants, we get
\begin{widetext}
	\begin{equation}\label{eq:drho_a}
		\begin{aligned}
			\langle \Delta \rho_1 \cdots \Delta \rho_n \rangle &= \nu \Delta t \int \sum_{k_1=1}^\infty \f{1}{k_1!} \pnt{ - \sum_{\alpha_1=1}^d \ell_{\alpha_1} \p_{1 \alpha_1} }^{k_1} \cdots \sum_{k_n=1}^\infty \f{1}{k_n!} \pnt{ - \sum_{\alpha_n=1}^d \ell_{\alpha_n} \p_{n \alpha_n} }^{k_n}
			\\
			& \quad \times \sum_{i=1}^N \rho_i (\bx_1, t) \cdots \rho_i (\bx_n, t) \cW(\bll) \dd \bll + \cO \pnt{ \Delta t^2 } .
		\end{aligned}
	\end{equation}
We use the properties of the Dirac delta function to simplify the last sum in Eq.~\eqref{eq:drho_a} as
\begin{equation}
	\begin{aligned}
		\sum_{i=1}^N \rho_i (\bx_1, t) \cdots \rho_i (\bx_n, t) = \delta (\bx_1 - \bx_2) \cdots \delta (\bx_{n-1} - \bx_n ) \rho (\bx_1, t) ,
	\end{aligned}
\end{equation}
yielding
\begin{equation}\label{eq:drho_g}
	\begin{aligned}
		\avg{ \Delta \rho_1 \cdots \Delta \rho_n } &= \nu \Delta t \int \brt{ \prod_{j=1}^n \sum_{k=1}^\infty \f{ \pnt{ - \bll \cdot \nabla_j }^{k} }{k!} } \rho \pnt{ \bx_1, t } \delta(\bx_1 - \bx_2) \cdots  \delta(\bx_{n-1} - \bx_n) \cW(\bll) \dd \bll + \cO \pnt{ \Delta t^2 }
		\\
		&= \nu \Delta t \int \brt{ \prod_{j=1}^n \pnt{ \ee^{ -\bll \cdot \nabla_j } - 1 } }  \rho \pnt{ \bx_1, t } \delta(\bx_1 - \bx_2) \cdots  \delta(\bx_{n-1} - \bx_n) \cW(\bll) \dd \bll + \cO \pnt{ \Delta t^2 } .
	\end{aligned}
\end{equation}
When substituting Eq.~\eqref{eq:drho_g} in Eq.~\eqref{eq:action}, a term of the following form appears
\begin{equation}
	\begin{aligned}
		\int_{\bx^n} \bar \rho (\bx_1, t) \pnt{ \ee^{ - \bll \cdot \nabla_1 } - 1 } & \brt{ \rho (\bx_1, t) \prod_{j=2}^n \bar \rho (\bx_j, t) \pnt{ \ee^{ - \bll \cdot \nabla_j } -1 } \delta ( \bx_1 - \bx_j ) }
		\\
		&= \int_{\bx^n} \rho (\bx_1, t) \brt{ \prod_{j=2}^n \delta ( \bx_1 - \bx_j ) \pnt{ \ee^{ \bll \cdot \nabla_j } -1 } \bar \rho (\bx_j, t) } \pnt{ \ee^{ \bll \cdot \nabla_1 } -1 } \bar \rho (\bx_1, t)
		\\
		&= \int_\bx \rho (\bx, t) \brt{ \pnt{ \ee^{ \bll \cdot \nabla_\bx } -1 } \bar \rho (\bx, t) }^n ,
	\end{aligned}
\end{equation}
\end{widetext}
where we have integrated by parts with respect to $\bx^n$ to get the second line, and we have integrated over $ \bx^{n-1} = \cur{ \bx_2, \dots, \bx_n } $ to obtain the third one. The dynamic action follows as
\begin{equation}
	\begin{aligned}
		\cA &= \int_{\bx, t} \brt{ \bar \rho \p_t \rho + \f{1}{\gamma} \p_\bx \bar \rho \cdot \int_\by \rho(\bx,t) \p_\bx v(\bx-\by) \rho(\by, t) }
		\\
		& \quad - \nu \int_{\bx, \bll, t} \rho (\bx, t) \sum_{n=1}^\infty \f{1}{n!} \brt{ \pnt{ \ee^{ \bll \cdot \nabla_\bx } - 1 } \bar \rho (\bx, t) }^n \cW(\bll)
		\\
		&= \int_{\bx, t} \brt{ \bar \rho \p_t \rho + \f{1}{\gamma} \p_\bx \bar \rho \cdot \int_\by \rho(\bx,t) \p_\bx v(\bx-\by) \rho(\by, t) }
		\\
		& \quad - \nu \int_{\bx, \bll, t} \rho (\bx, t) \cur{ \exp \brt{ \pnt{ \ee^{ \bll \cdot \nabla_\bx } - 1 } \bar \rho (\bx, t) } - 1 } \cW(\bll) .
	\end{aligned}
\end{equation}
The linear order in the conjugated field gives back the Fokker-Planck equation, so that the dynamic action can be expressed
\begin{equation}
	\begin{aligned}
		\cA &= \int_{\bx, t} \pnt{ \bar \rho \p_t \rho + \p_\bx \bar \rho \cdot {\bf D}_\bx \rho }
		\\
		& \quad + \f{1}{\gamma} \int_{\bx, \by, t} (\rho \p_\bx \bar \rho) (\bx,t) \p_\bx v(\bx-\by) \rho(\by,t)
		\\
		& \quad + \text{higher orders in } \bar\rho ,
	\end{aligned}
\end{equation}
where we have used the representation of $\bf D_\bx$ in term of the jump distribution in~\eqref{eq:L}. We introduce a new set of fields $\cur{ a, \bar a}$ related to the previous one $\cur{ \rho, \bar \rho}$ through the Cole-Hopf transformation as
\begin{equation}
\bar a = \ee^{\bar \rho} , \quad a = \rho \ee^{-\bar \rho} .
\end{equation}
Our aim is to show that the dynamic action can be simplified as
\begin{equation}
	\begin{aligned}
		\cA &= \int_{\bx, t} \pnt{ \bar a \p_t a + \p_\bx \bar a \cdot {\bf D}_\bx a }
		\\
		&\quad + \f{1}{\gamma} \int_{t,\bx,\by} \pnt{ a \p_\bx \bar a } (\bx, t) \cdot \p_\bx v (\bx-\by) (a \bar a) (\by, t)
		\\
		&= \int_{\bx, t} \brt{ \bar a \p_t a + a \p_\bx \bar a \cdot \int_\by \p_\bx v (\bx-\by) (a \bar a) (\by, t) }
		\\
		& \quad - \nu \int_{\bx, \bll, t} p(\bll) \brt{ \bar a(\bx+\bll,t) - \bar a(\bx, t) } a(\bx, t) ,
	\end{aligned}
\end{equation}
which is equivalent to demonstrate that
\begin{equation}\label{eq:eq_CH}
	\begin{aligned}
		\int_{\bx} \ee^{\bar \rho} & \pnt{ \ee^{ - \bll \cdot \nabla_\bx } - 1 } \pnt{ \rho \ee^{- \bar \rho} }
		\\
		&= \int_{\bx} \rho \cur{ \exp \brt{ \pnt{ \ee^{ \bll \cdot \nabla_\bx } - 1 } \bar \rho } - 1 }
	\end{aligned}
\end{equation}
The operator $\ee^{ \bll \cdot \nabla_\bx }$ corresponds to the translation operator by a vector $\bll$, which effect on an arbitrary function $f(\bx)$ is given by
\begin{equation}
\ee^{ \bll \cdot \nabla_\bx } f(\bx) = f( \bx + \bll ) .
\end{equation}
Therefore, Eq.~\eqref{eq:eq_CH} can be written as
\begin{equation}
	\begin{aligned}
		\int_{\bx} & \brt{ \rho ( \bx - \bll ) \ee^{ \bar \rho (\bx) - \bar \rho  ( \bx - \bll ) } - \rho (\bx) }
		\\
		&= \int_{\bx} \rho (\bx) \brt{ \ee^{ \bar \rho ( \bx + \bll ) - \bar \rho (\bx)  } - 1 } .
	\end{aligned}
\end{equation}
Eventually, by translating the argument as $\bx \to \bx - \bll$ in the first term of the rhs, the validity of this equation follows immediately.

% ===============================================================================

\bibliography{nongaussian-bibliography}

\end{document}